\documentclass[journal]{IEEEtran}

\usepackage{caption, subcaption, amsmath, url, pifont, graphicx, cite, multirow, rotating, array, mathtools, bm, threeparttable, booktabs, xcolor, amssymb, tikz, lineno, amsfonts, setspace}
\usepackage[hidelinks]{hyperref}
\usepackage[outdir=./]{epstopdf}

\makeatletter
\def\endthebibliography{
	\def\@noitemerr{\@latex@warning{Empty `thebibliography' environment}}
	\endlist
}
\makeatother

\definecolor{lime}{HTML}{A6CE39}
\DeclareRobustCommand{\orcidicon}{
\begin{tikzpicture}
\draw[lime, fill=lime] (0,0)
circle[radius=0.16]
node[white]{{\fontfamily{qag}\selectfont \tiny \.{I}D}};
\end{tikzpicture}
\hspace{-2mm}
}
\foreach \x in {A, ..., Z}{%
\expandafter\xdef\csname orcid\x\endcsname{\noexpand\href{https://orcid.org/\csname orcidauthor\x\endcsname}{\noexpand\orcidicon}}
}

\begin{document}

\title{A color temperature-based high-speed decolorization: an empirical approach for tone mapping applications}
\author{Prasoon~Ambalathankandy\hspace{-1.5mm}\orcidP{}, Yafei~Ou\hspace{-1.5mm}\orcidO{}, and~Masayuki~Ikebe\hspace{-1.5mm}\orcidI{}
	
\thanks{Prasoon Ambalathankandy and Yafei Ou are with the Research Center For Integrated Quantum Electronics, Hokkaido University, Sapporo 060-0813, Japan, and also with the Graduate School of Information Science and Technology, Hokkaido University, Sapporo 060-0814, Japan.}
\thanks{Masayuki Ikebe is with the Research Center For Integrated Quantum Electronics, Hokkaido University, Sapporo 060-0813, Japan (e-mail: ikebe@ist.hokudai.ac.jp).}
}

{}
\bibliographystyle{IEEEtranTIE}

\maketitle

\begin{abstract}
Grayscale images are fundamental to many image processing applications like data compression, feature extraction, printing and tone mapping. However, some image information is lost when converting from color to grayscale. In this paper, we propose a light-weight and high-speed image decolorization method based on human perception of color temperatures. Chromatic aberration results from differential refraction of light depending on its wavelength. It causes some rays corresponding to cooler colors (like blue, green) to converge before the warmer colors (like red, orange). This phenomena creates a perception of warm colors ``advancing'' toward the eye, while the cool colors to be ``receding'' away. In this proposed color to gray conversion model, we implement a weighted blending function to combine red (perceived warm) and blue (perceived cool) channel. Our main contribution is threefold: First, we implement a high-speed color processing method using exact pixel by pixel processing, and we report a $5.7\times$ speed up when compared to other new algorithms. Second, our optimal color conversion method produces luminance in images that are comparable to other state of the art methods which we quantified using the objective metrics (E-score and C2G-SSIM) and a subjective user study. Third, we demonstrate that an effective luminance distribution can be achieved using our algorithm by using global and local tone mapping applications.

\end{abstract}
	

\section{Introduction}
\label{sect:intro}
Grayscale channels, which reflect image luminance, are used for various applications such as printing, tone mapping, data compression, and feature extraction. Thus, obtaining luminance along with human perception has a key role for decolorization, which converts RGB channels to high-quality gray ones. For example, High Dynamic Range (HDR) compression is ideally performed by tone mapping the luminance channel for the lower computational and memory cost. However, applying well-known luminance channels such as Y of YCbCr or V of HSV does not guarantee appropriate tone mapping, as these channels do not reflect human perceptions. Therefore, decolorization has gathered considerable attention and various sophisticated methods to achieve perceptual decolorization have recently been proposed. These methods can be classified into global and local methods. Global methods can define only one conversion function for all pixels, and most of these methods use all pixels in the image to determine the function. On the other hand, local ones process the target from neighboring pixels in the same way as a spatial filter, the function is different for each pixel. However, both types of methods face the issue of calculation cost, which comes from optimization iterations or spatial filter processing. 

We have developed a fast decolorization method that reflects the perception of warm and cool colors which is well known in psychophysics studies\cite{morton1998color}. Colors are arranged according to their wavelengths on the color wheel, the ones with longest wavelengths are on the right side of the wheel and are known as warm colors, as they evoke warmth. These hues include shades of red, yellow, and orange. On the other hand, green, blue and violet which have shorter wavelengths are placed on the left side of the color wheel, and are perceived as cool colors. The color of an object in a scene affects our perception of its apparent depth and this phenomenon has been exploited by many artists. This optical illusion has been studied by psychologists, and early researchers explored color-depth relationship. One of the widely accepted theory explains this phenomenon is due to fact that shorter wavelengths of visible light are refracted more than longer wavelengths \cite{sundet1978effects}. In other words, an equidistant source of different wavelengths cannot be focused simultaneously onto our retina. This phenomena is called as chromatic aberration and we discuss it in detail in section \ref{ca}. In our decolorization method, we combine the idea of warm/cool colors with the Helmholtz-Kohlrausch effect \cite{Nayatani1}. On that account, we make two assumptions, which are: $(i)$ warm colors (mainly including $R$) are lighter than $Y$ of YCbCr and $(ii)$ mixed colors are darker than the $Y$ or $L$ of CIE with the same luminance. To satisfy these assumptions, we use a weighted blending of RGB channels and remap them to warm/cool colors on the luminance channel. Following are our main contributions:

\begin{itemize}
	\item We achieve high-speed color mapping by exact pixel-by-pixel processing. 
\end{itemize}	

\begin{itemize}
	\item We obtain luminance comparable to that of optimization-based methods.  
\end{itemize}	

\begin{itemize}
	\item We demonstrate effective luminance distribution by performing objective evaluations and a subjective user study. 
\end{itemize}	

\begin{figure*}[t]
	\centering
	\includegraphics[width = \textwidth]{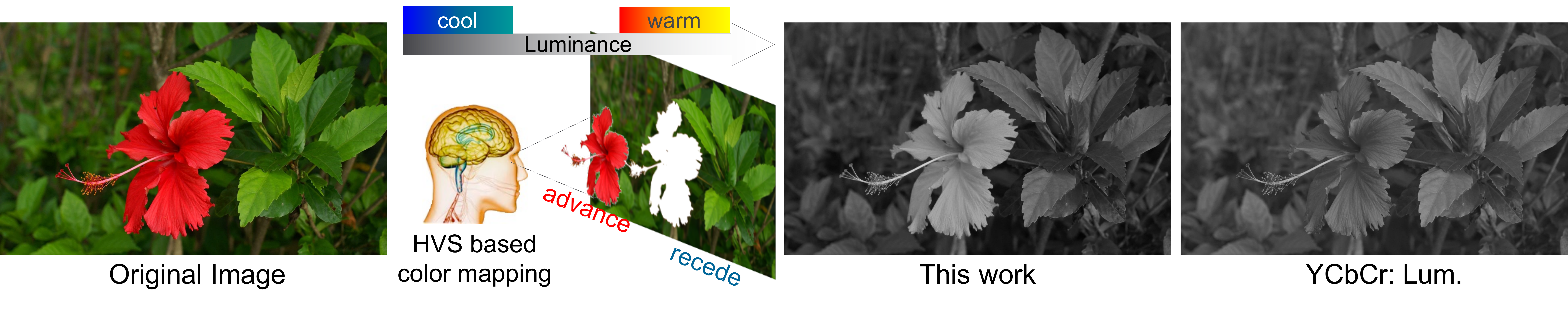}
	\caption{Main concept of our decolorization method: Human perception of warm and cool colors. Warm colors ``advance'' toward the eye, while cool colors ``recede''. In this work we are able to accurately reflect the human perception of color temperatures, whereas this phenomenon is non-existent in conventional YCbCr color space.}   
	\label{fig:1}
\end{figure*}

\begin{figure}[!t]
	\centering
	\includegraphics[width = \linewidth]{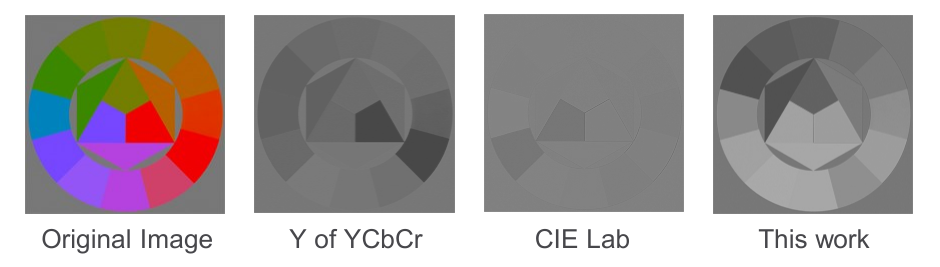}
	\caption{Comparison of luminance components obtained using YCbCr, CIELAB and our proposed method.}
	\label{fig:2}
\end{figure}

There are many well defined methods to convert any color image to a grayscale image. An effortless procedure is to assign different weights to color channels, in order to have the same luminance in the grayscale image as the original color image. For example, in the MATLAB function \texttt{rgb2gray}, it converts any RGB values to grayscale ($Gray$) values by forming a weighted sum of the $R$, $G$, and $B$ components as $Gray = 0.2989\times R + 0.5870\times G + 0.1140\times B$. This function operates under an assumption that human visual system is more sensitive to green color. When operating with CIELab and YUV color spaces, one could directly obtain luminance channel as the grayscale version of the color image as they consider the luminance and color channel to be independent. But, such crude approaches will fail to preserve image contrast as shown in these examples (see Fig. \ref{fig:1} and Fig. \ref{fig:2}).

In several real-world image/video processing applications like detail enhancement, image matching, and segmentation under different illumintion a 1-D grayscale image has to be obtained from its corresponding 3-D color image. However, mapping the 3-D color information onto a 1-D grayscale image while retaining the original contrast and fine details is a challenging problem. Additionally, implementing decolorization algorithms with a reasonable computational efficiency is pivotal for realising their real-time applications. Many studies have been carried out to develop novel decolorization methods. These mapping methods can be categorized into global 
\cite{gooch2005color2gray,grundland2007decolorize,smith2008apparent,kim2009robust,lu2012contrast} and local methods \cite{ancuti2010image,ancuti2016laplacian,liu2015gcsdecolor,song20161}. In local mapping methods, the same color pixel within an image could be mapped into different grayscale values depending on its spatial location. Ideally this undesirable as such output images may be perceived as unnatural. On the other hand, in global 
mapping methods same color pixels within an image irrespective of its spatial location are mapped to same grayscale values. Thus, global methods are more likely to produce grayscale images that are perceived to appear natural. 

In the global methods category, Gooch et al. \cite{gooch2005color2gray} proposed a global decolorization algorithm that can be implemented by solving the optimization problem for all image pixels. Then, Kim et al. \cite{kim2009robust} aimed at high-speed processing by simplifying Gooch's method. Smith et al. \cite{smith2008apparent} used unsharp masking and the Helmholtz-Kohlrausch effect (H-K effect) model of Nayatani et al. \cite{Nayatani1}. Nayatani's model \cite{Nayatani1} is merely an experimental model for the effect of the CIELUV chrominance component for human perception. On the other hand, the method of Lu et al. \cite{lu2012contrast} is focused on converting a color image into a gray image with high contrast. The main advantage of these methods is transformation consistency, i.e., the same color is converted to the same grayscale. However, speed remains a problem for  these methods. Most of the local methods are aimed at speeding up the method of Lu et al \cite{lu2012contrast}. To enhance image contrast, Ancuti et al. \cite{ancuti2016laplacian} adopted the strategy of using a Laplacian filter and Song et al. \cite{song20161} used a Sobel filter. Although local methods are effective in terms of contrast emphasis, they are disadvantageous in terms of tone mapping because 
conversion consistency is not maintained and it differs from human perception.

Recently, some machine learning-based techniques have also been proposed for image decolorization  \cite{liu2019image,cai2018perception,hou2017deep,lin2008learning,zhang2018contrast}. Cai et al. \cite{cai2018perception} proposed a method, that used the perceptual loss function to pretrain VGG-Net \cite{simonyan2018very}. However, it is difficult to control and many of their output images are far from human perception additionally, the computational cost are high. Processing an image of $256 \times 256$ size it requires roughly 30 seconds on a single Nvidia GeForceGTX 1080 GPU. Zhang et al., proposed a CNN framework that combines local and global image features \cite{zhang2018contrast}. However, their network framework do not account for exposure features \cite{liu2019image}. Lin et al.'s method \cite{lin2008learning} by utilizing a database of 50 images from the Corel dataset produced 50 grayscale images using the Color2Gray algorithm \cite{gooch2005color2gray}. With these 50 input/output image pairs as training examples for their partial differential equations-based (PDE)learning system, the learn color2gray mapping. The PDE system produced images of comparable quality to \cite{gooch2005color2gray}. However, for an input image of size $n \times n$ their PDE color mapping algorithm's computational complexity is $\textit{O}(n^{2})$. Under these circumstances, a high-speed method for 
generating grayscale images that accurately captures human perception has not yet been developed. To make tone maps valid, it is necessary to develop such a method.

\section{Proposed Method}

\subsection{Problem Definition}
Luminance components such as $Y$ of YCbCr and CIE $L$ have been used in various image processing applications; however, they do not accurately reflect human perception (Fig.\ref{fig:1} and Fig. \ref{fig:2}). Figure \ref{fig:1}  shows how the warm colored flower (red) advances towards the eye of the observer, while the background mainly green recedes. Using the luminance channel of the conventional YCbCr color space this phenomenon is absent. But, in our method we are able to capture the warmth $R$ (red) component as perceived in human perception. Figure \ref{fig:2} shows that the $R$ (red) and $B$ (blue) components do not come even close to the perception in the luminance component of CIELAB. Moreover, mixed color components such as mud yellow tend to appear dark for people. In this study, we conceived the idea that RGB weight functions for alpha blending can reproduce this phenomenon.

\subsection{Luminance mapping using red and blue weighting function}

\begin{figure}[!t]
	\centering
	\includegraphics[width = \linewidth]{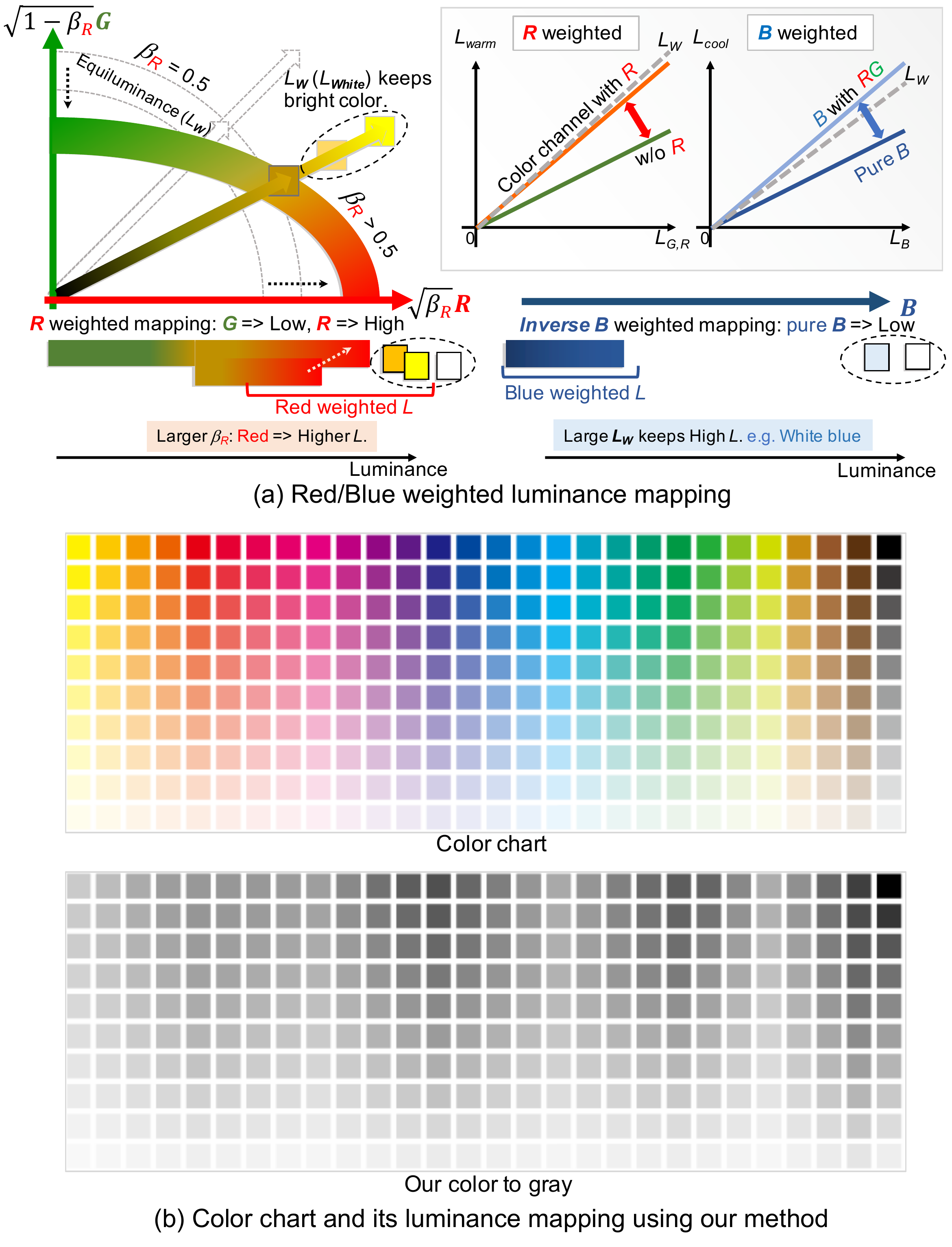}
	\caption{(a) Remodeling of luminance space based on our two weighting functions one each for warm and cool colors. Here, luminance is defined as the Euclidean distance of warm/cool color. (b) Decolorization of a color chart illustrating our method's effectiveness in generating brighter shades of gray for warm colors.}
	\label{fig:3}
\end{figure}

We considered the idea of color mapping by combining the warm/cool color and the Helmholtz-Kohlrausch \cite{Nayatani1}. Psychophysical studies find that, warm and cool colors impact our visual perception of the objects that we see. For example, the red color associated with fire/sun advances toward the eye, creates an illusion of heat and therefore perceived as warmth and comforting. On the other hand, cool colors have reverse effects of warm colors. Receding from the eye of the observer, cool colors reminds of the earthy objects, like meadows and oceans. These hues often are perceived as cool and refreshing \cite{morton1998color}\cite{ou2004study}. In our decolorization method, we developed two weighting functions as shown in Fig \ref{fig:3}. One function for remapping warm colors and the other for remapping cool colors. In our method, actual luminance is defined as the Euclidean distance of weighted warm/cool luminance including the $W$ (white) channel. Essential luminance is given by 

\begin{footnotesize}
	\begin{equation}
		L_{WHITE}= \sqrt{\frac{R^{2}+G^{2}+B^{2}}{3}}
		\label{eq:Lw1}
	\end{equation}
	\begin{equation}
		L_{B} = B
		\label{eq:Lb}
	\end{equation}
	\begin{equation}
		L_{G,R}= \sqrt{\beta_{R}R^{2}+(1-\beta_{R})G^{2}}, \; (0.5 < \beta_{R} < 1)   
		\label{eq:Lgr}
	\end{equation}
\end{footnotesize}
As we know, $L_{WHITE}:W$ is the Euclidean distance of $RGB$ channels; $L_{B}$ is the $B$ channel as it is. The warm color function is also defined as the Euclidean distance by $\beta_{R}$ weighted $R$ and $G$(green) vectors. 
The $\beta_{R}$ determines the bias of $R$ components in the replacement; when the $\beta_{R}$ becomes higher, more $R$ components than $G$ components are rated. The remappings made by the blending function using the $R$ and $B$ 
ratio of $RGB$ values are given by 

\begin{footnotesize}
	\begin{equation}
		L_{WARM} = \frac{R}{R+G+B}\cdot L_{G,R} + \left(1-\frac{R}{R+G+B}\right)\cdot L_{WHITE}
		\label{eq:Lwarm}
	\end{equation}
	\begin{equation}
		L_{COOL} = \left(1-\frac{B}{R+G+B}\right)\cdot L_{B} + \frac{B}{R+G+B}\cdot L_{WHITE}
		\label{eq:Lcool}
	\end{equation}
\end{footnotesize}

\begin{figure}[!t]
	\centering
	\includegraphics[width = \linewidth]{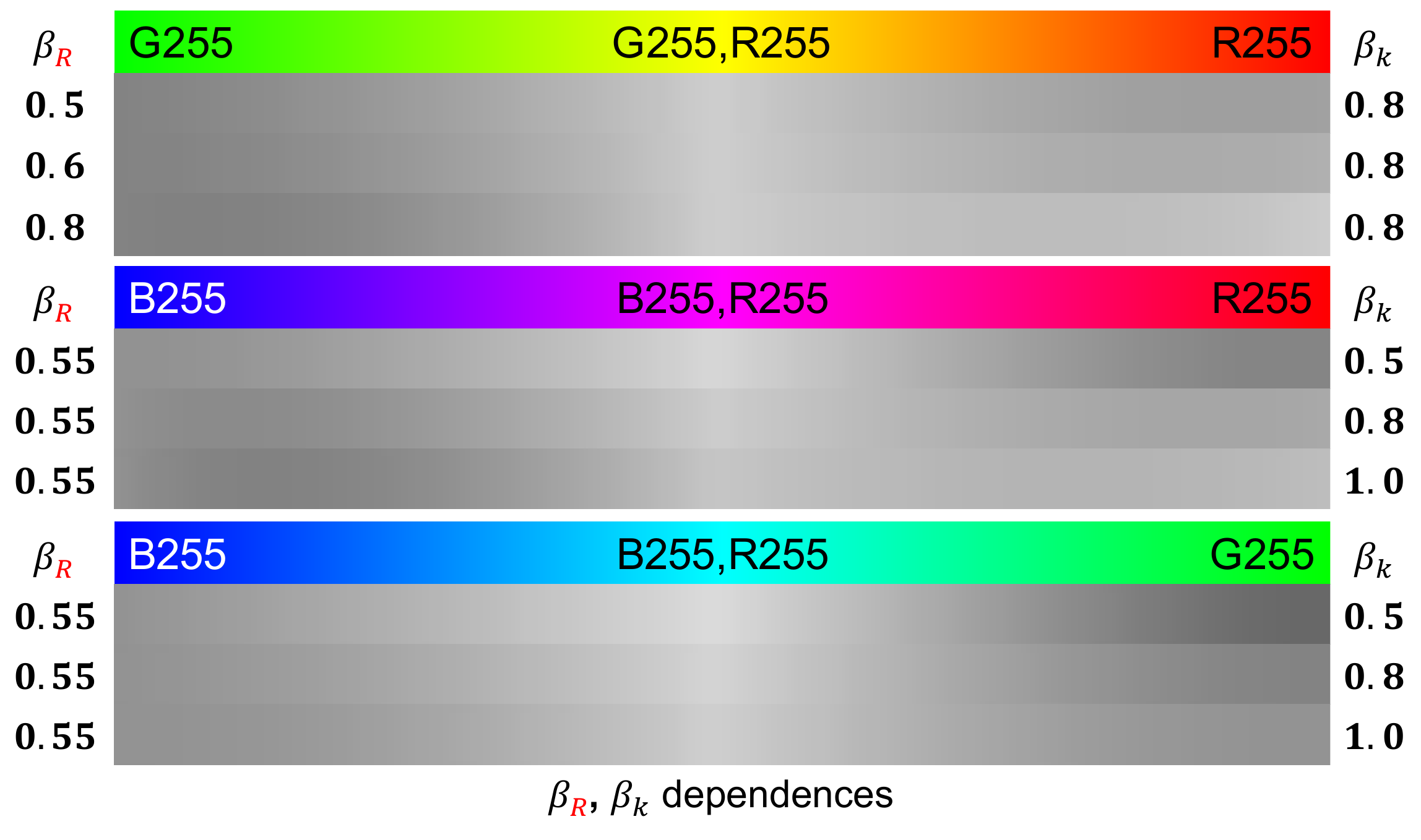}
	\caption{Impact of $\beta_{R}$ and $\beta_{k}$ on output luminance value. (top) $\beta_{k}$ set to 0.8 and for changing $\beta_{R}$ we move from $G_{max}$ to $R_{max}$. (middle) $\beta_{R}$ set to 0.55 and for changing $\beta_{k}$ we move from $B_{max}$ to $R_{max}$. (bottom) $\beta_{R}$ set to 0.55 and for changing $\beta_{k}$ we move from $B_{max}$ to $G_{max}$}
	\label{fig:4}
\end{figure}

\begin{figure*}[!t]
	\centering
	\includegraphics[width = \textwidth]{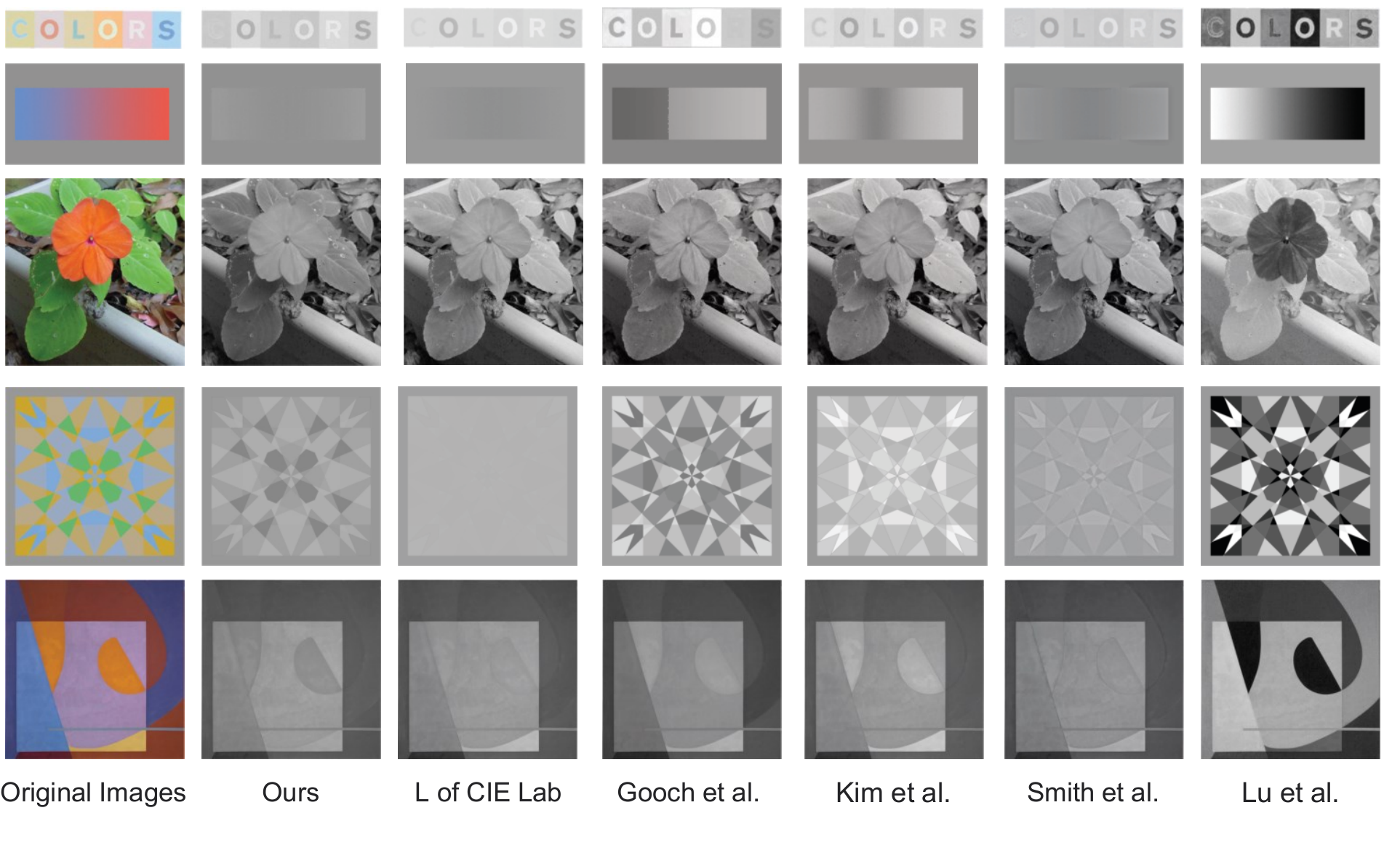}
	\caption{Color to gray conversion comparison with other decolorization methods.}
	\label{fig:5}
\end{figure*}

In the $L_{WARM}$, using $R$ weighting, we can map $\frac{G}{R}$ to $low/high$ luminance separately. In Inverse $B$ weighting, pure $B$ components are assigned to low luminance in the $L_{COOL}$ . Since both functions are 
blended with $L_{WHITE}$, bright orange/yellow and sky blue, which include high white components, are mapped to higher luminance. Finally, we obtain the luminance channel $L$, which is given by 

\begin{footnotesize}
	\begin{equation}
		L= \sqrt{\beta_{k}L^{2}_{WARM}+(1-\beta_{k})L^{2}_{COOL}} \; (0.5 < \beta_{k} < 1)  
		\label{eq:Lw2}
	\end{equation}
\end{footnotesize}
In this study, we mainly used warm-color weighting luminance in experiments and set the $\beta_{k}$ higher. We set two parameters relating to color component emphasis as follows: $(\beta_{R} = 0.55; \beta_{k} = 0.8)$, and from Fig \ref{fig:4} it can be easily understood how $\beta_{R}$ and $\beta_{k}$ can bias the resulting luminance values.

\subsection{Evaluation of our algorithm implementation}

\begin{figure}[!t]
	\centering
	\includegraphics[width = \linewidth]{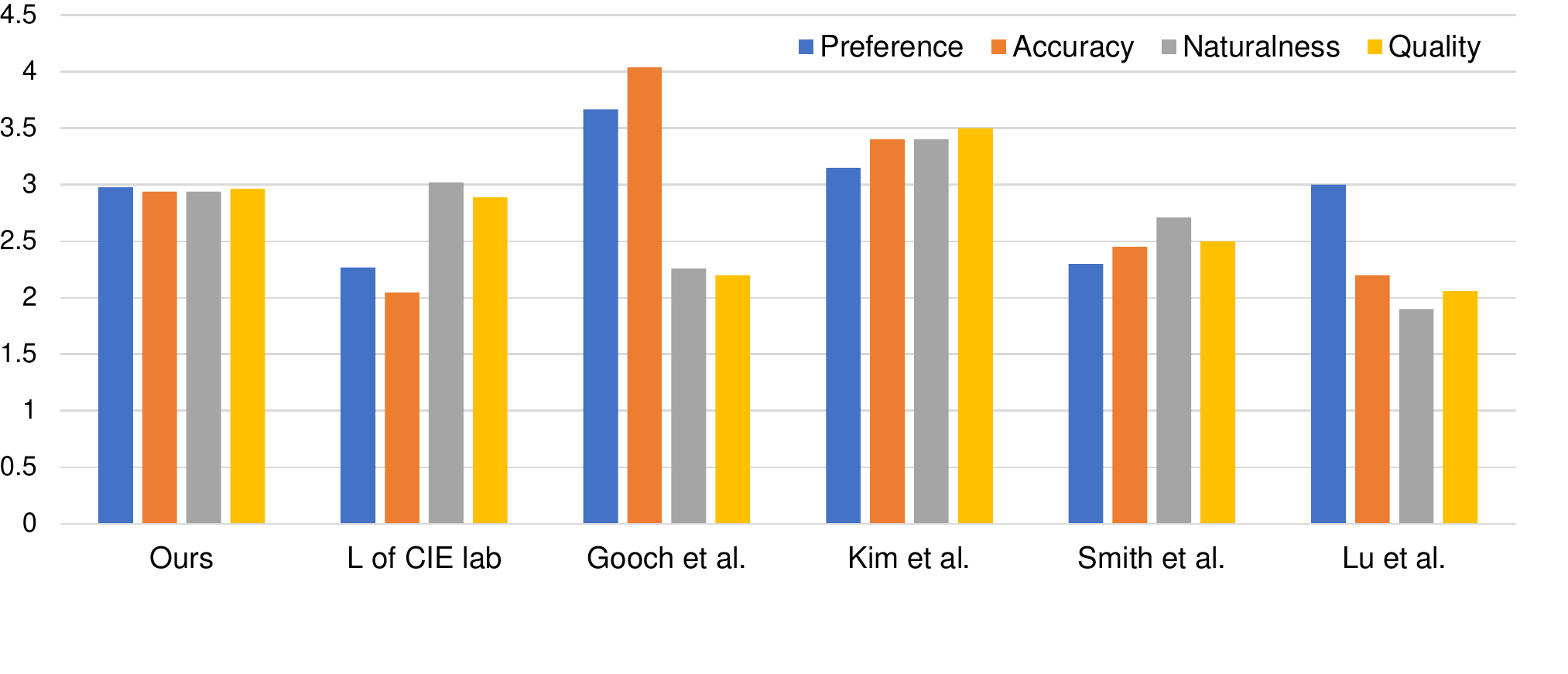}
	\caption{Subjective visual test survey and mean opinion score}
	\label{fig:vt}
\end{figure}

When evaluating our method, we focused on the following points:
 \begin{itemize}
 	\item Visual comparison with other decolorization methods 
 	\item Compare their processing speeds. 
\end{itemize}	
 	The evaluations demonstrated that our method does reflect human perception better than or equal to other optimized methods (see perceptual test). It delivers high-speed processing, and is a useful tool for many image processing applications. Figure \ref{fig:5} shows a comparison between our proposed method and other decolorization methods, and the subjective evaluation is shown in Fig. \ref{fig:vt}. For this comparison we have used images from Cadik’s dataset \cite{cadik2008perceptual}. In Fig. \ref{fig:5}  first and second columns from the left are original images and images obtained with our method respectively. The images in the third column from the left were obtained with the $L$ component of the CIELAB color space, which is a reversible model reflecting human visual characteristics. The images in the fourth to sixth columns were obtained from global decolorization methods, which means only one conversion function is applied to each pixel. For example, in these methods, optimization techniques referring to whole pixels are applied to luminance conversion without regard to the brightness perceived by human perception. The images in the far right column were obtained with a local method that refers to pixel values in the local patch of the image for contrast enhancement.

\begin{figure*}[!t]
	\centering
	\includegraphics[width = \textwidth]{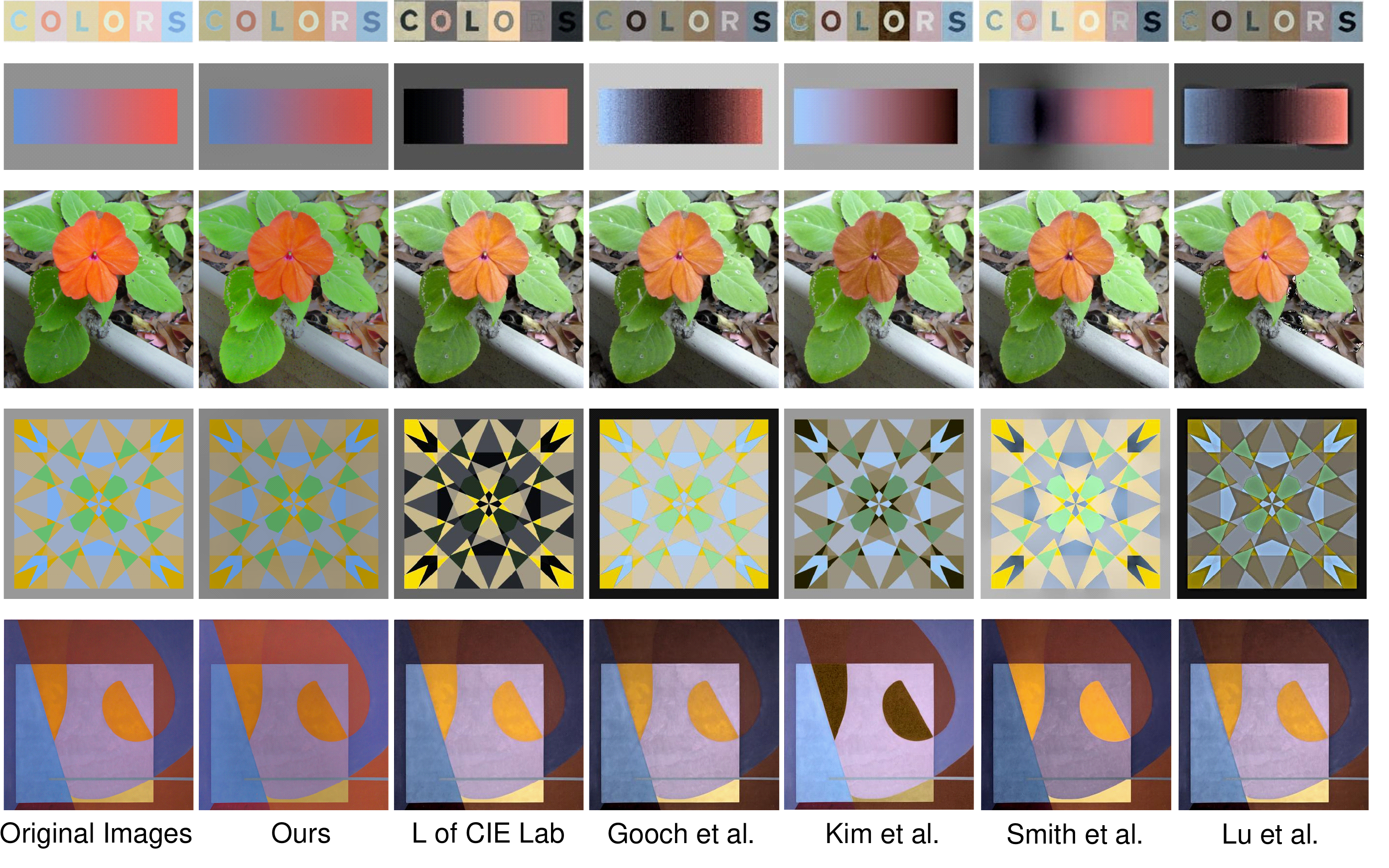}
	\caption{Effects of decolorization techniques on tone mapped images that are generated using Shan et al.'s tone mapping operator \cite{shan2009globally}.}
	\label{fig:tmo_1}
\end{figure*}

In the $L$ component of the CIELAB color space, the base luminance is $Y$ in the YCbCr color space; the luminance is also mapped along with the color order of $Y$. In the image (third column, fourth row), since the luminance of all colors are the same in $Y$, output values are also the same in the CIELAB color space. However, we can perceive the contrast in this image. Thus, an appropriate conversion is required; our method can generate the perceived contrast in this image. Gooch et al. \cite{gooch2005color2gray} obtained the highest average C2G-SSIM score but clearly a step artifact occurs in the gradation image (fourth column, second row). Kim et al. \cite{kim2009robust} proposed an improved version of Gooch's method that achieves high-speed optimization and reflects the Helmholtz-Kohlrausch \cite{Nayatani1}. In this method, since a weighting function is applied for expanding luminance distribution of whole pixel colors in the image along with chrominance, its conversion becomes different in each image. Thus, over-enhancement is observed in the images (fifth column, first row) and (fifth column, fourth row). The output images 
of our method are similar to those of Smith et al. \cite{smith2008apparent}. Their method also uses the Helmholtz-Kohlrausch, but it also requires a lot of processing time for post-unsharp-mask filtering. In Lu et al. \cite{lu2012contrast}, their method does not reflect human perception; they try generating high contrast images for mask images that are input to an edge-preserving filter such as a guided filter. 

\begin{table*}[!t]\scriptsize
	\renewcommand{\arraystretch}{1.2}
	\caption{Run-time comparison table with other decolorization algorithms.}
		\centering
	\label{table:1}
	\begin{tabular}{|p{6em}|p{6.5em}|p{7em}|p{7.5em}|p{5em}|p{6.5em}|p{8em}|}
		\hline
		Algorithm &  Processing time  & Image size W$\times$H & CPU Clock Speed &  Optimization &  Process & Normalized Time$^\star$ \\ \hline
		Gooch et al. \cite{gooch2005color2gray}   &  25.7s & 200$\times$200    & -GPU- & \checkmark &Global & N/A\\ \hline % O(4)
		Kim et al.  \cite{kim2009robust} &  102ms & 320$\times$240    &  2.66GHz & \checkmark & Global & 1.30$\mu$s \\ \hline % O(4)
		Smith et al. \cite{smith2008apparent}   &  6.7s & 570$\times$593    & 3.0GHz &$\times$ &Global+Local & 22.02$\mu$s \\ \hline % O(4)
		Lu et al.  \cite{lu2012contrast} &  800ms & 600$\times$600    &  3.80GHz & \checkmark &Global Contrast & 3.12$\mu$s \\ \hline % O(4)
		Song et al.  \cite{song20161} &  40ms & 320$\times$240    & N/A & $\times$ &Local Contrast&N/A\\ \hline % O(4) et al.   &  800    & 600$\times$600    &  3.80(GHz) &\checkmark &Local Contrast\\ \hline % O(4)
		Ancuti et al. \cite{ancuti2016laplacian}  &  100ms & 800$\times$600   & 2.5GHz & $\times$ &Local Contrast & 0.19$\mu$s \\ \hline %
		L of CIE Lab    &  25.57ms & 800$\times$600    &2.7GHz &$\times$ & Global & 0.053$\mu$s \\ \hline
		Ours low res &   \textbf{16.71ms}    & 800$\times$600    & 2.7GHz &$\times$ &Global& \textbf{0.034$\mu$s}\\
		Ours high res &  \textbf{202.05ms}    & 3008$\times$2008   & 2.7GHz &$\times$ &Global& \textbf{0.033$\mu$s}\\
		\hline		
	\end{tabular}
	\flushleft
	$^\star$ Normalized time is the processing time normalized by frequency (2.7 GHz) and divided by the number of pixels. It indicates the effective processing time per pixel.
\end{table*}

For our perceptual evaluation we conducted a user study with two objectives: 
\begin{itemize}
	\item Evaluate the decolorization process by measuring accuracy and preference \cite{cadik2008perceptual,lu2014contrast}.
\end{itemize}
\begin{itemize}
	\item Assess the effectiveness of our decolorization method as an image pre-processing tool for tone mapping application.
\end{itemize}
Our study group of 15 students (9 males, 6 females, average age = 23) were shown twelve sets of images (refer Figs. \ref{fig:5} and \ref{fig:tmo_1} ). They were asked to evaluate Fig. \ref{fig:5} and assign points to these images, on a scale of 1 (low) to 5 (high) for accuracy and preference. For our first evaluation, to measure the accuracy they compared the original color images to their decolorized output image. For the preference measurements, the user group compared the decolorized images only, i.e., without the corresponding color image. In our next user study we measured the naturalness and overall quality of the tone mapped images. For tone mapping we chose a MATLAB algorithm of Shan et al. \cite{shan2009globally} as they can tonemap standard images and it can be performed by simple parameter settings. Color tone mapped images for the same set shown in Fig. \ref{fig:tmo_1} were evaluated for naturalness and perceptual quality, i.e., the users rated 5 for images that were perceived as natural and observed minimal artifacts. In our test setup we used a Microsoft Surface Pro 7 set to its native resolutions and the lighting of the test room was slightly dim. The compiled response of the volunteers are presented in Fig. \ref{fig:vt}. 

Table \ref{table:1} lists the processing speed of various methods, our proposed implementation and the CIELAB were implemented in C++; these codes were executed on an Intel Core i5-5257U (2.70GHz) CPU without any multicore, multithread or SIMD operations. The results confirmed that our method had the fastest runtime among the methods compared. It is worth noticing that it exceeded CIELAB in runtime; this indicates it also has advantages in total calculation cost including post-processing. Its computational complexity is only $\textit{O}(1)$ because it performs exact pixel by pixel processing, referring only to the RGB value at each pixel. Among other methods, the one developed by Gooch et al. was reported to have $\textit{O}(n^{4})$ computational complexity. Using $\textit{O}(1)$ spatial filtering it is possible to develop $\textit{O}(1)$ local methods \cite{ancuti2016laplacian,song20161}, but the filter calculations required would degrade their run-time in comparison to our decolorization method. The proposed color to gray technique has demonstrated faster run-time than local methods by maintaining global coherence, which means the conversions were the same in all pixels. 

\begin{figure}[t]
	\centering
	\includegraphics[width =\linewidth]{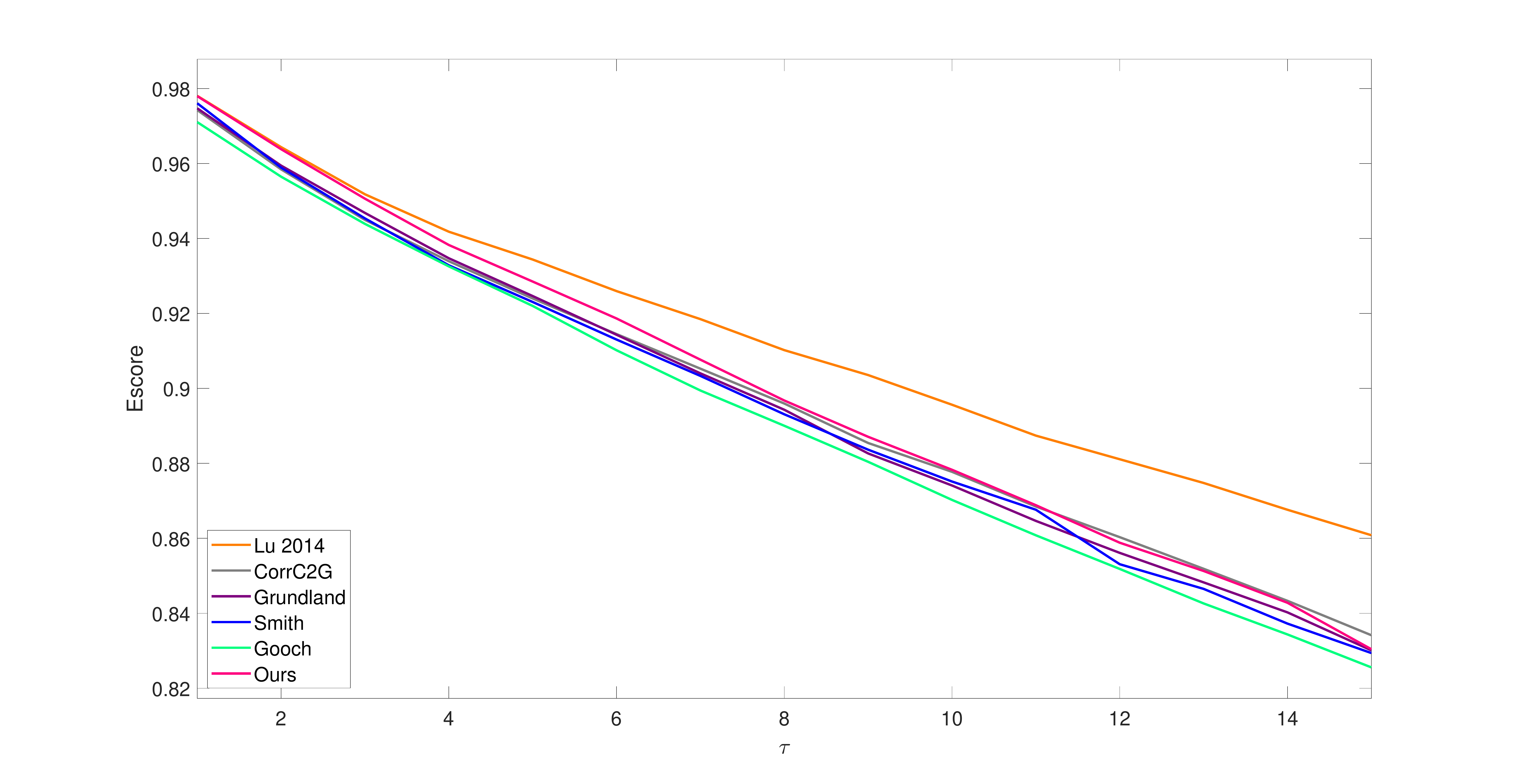}
	\caption{Comparison of six color to gray methods based on the E-score.}   
	\label{fig:plotE}
\end{figure}

\subsection{Experimental results}
In our experiments, we utilized the color250 dataset which comprises of 250 natural and synthetic color images \cite{lu2014contrast}. To quantitatively evaluate our decolorization algorithm we choose two objective metrics: E-score and C2G-SSIM by Ma et al \cite{ma2015objective}. 
E-score is a joint measure proposed by Lu et al., a harmonic mean which is computed by combining two metrics: Color Contrast Preserving Ratio (CCPR), and Color Content Fidelity Ratio (CCFR) \cite{lu2014contrast}. The CCPR is useful in maintaining the color contrast in decolorization images which is perceivable to humans. Specifically, when the color difference is smaller than a certain threshold value, it becomes undetectable to humans. Furthermore, CCFR estimates if the decolorization image is accurate in terms of structures when compared to the original color image. C2G-SSIM is new color to gray objective evaluation metric based on the Structural Similarity (SSIM) index quality metric \cite{wang2004image}. The C2G-SSIM generates quality map and has good correlation with HVS subjective preference. Table \ref{tab_om} presents the average E-score and C2G-SSIM for the color 250 dataset in comparison with other decolorization methods.

\begin{figure}[t]
	\centering
	\includegraphics[width = \linewidth]{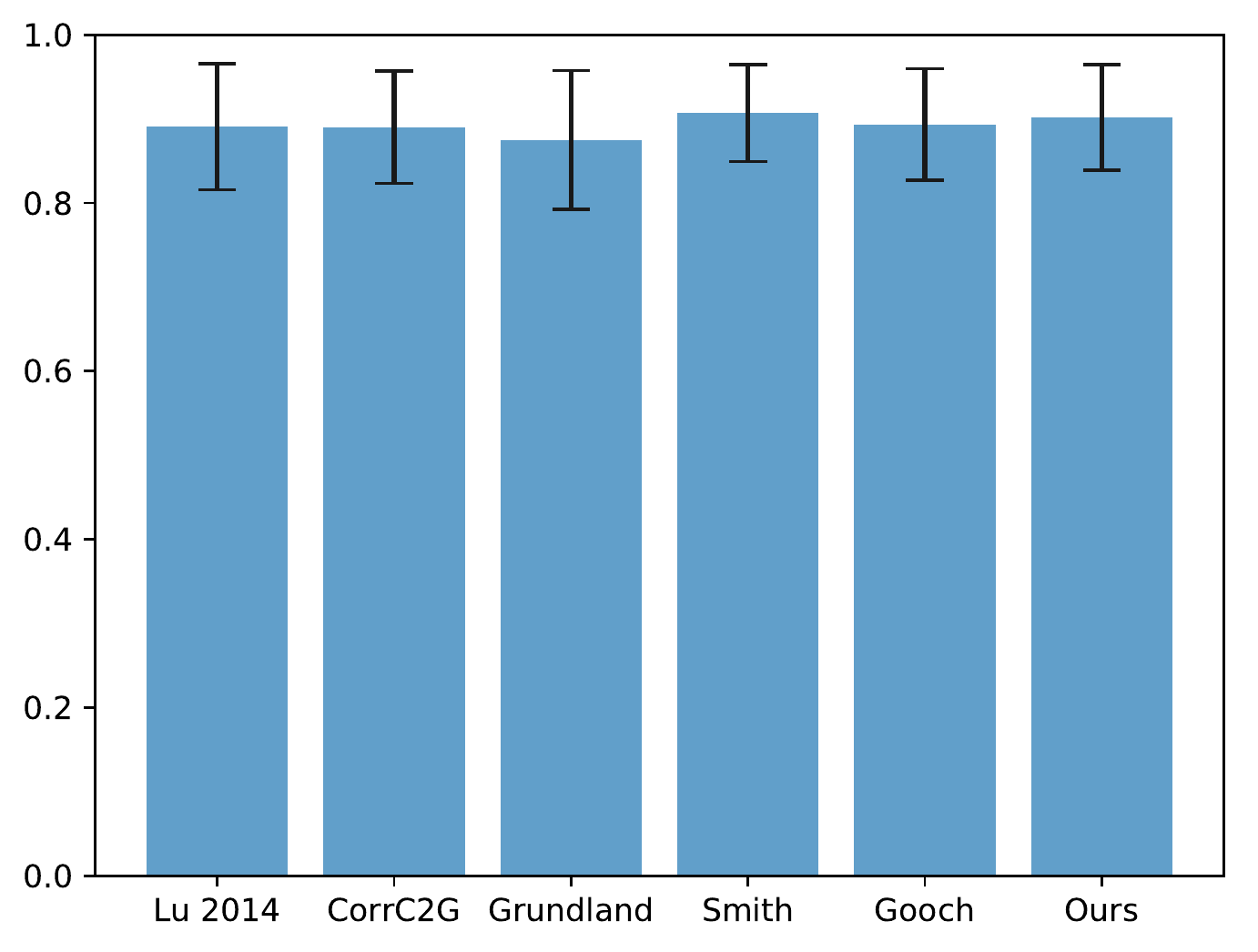}
	\caption{Comparison of six color to gray methods based on the C2G-SSIM score.}   
	\label{fig:ssim}
\end{figure}

\begin{figure*}[t]
	\centering
	\includegraphics[width = \textwidth]{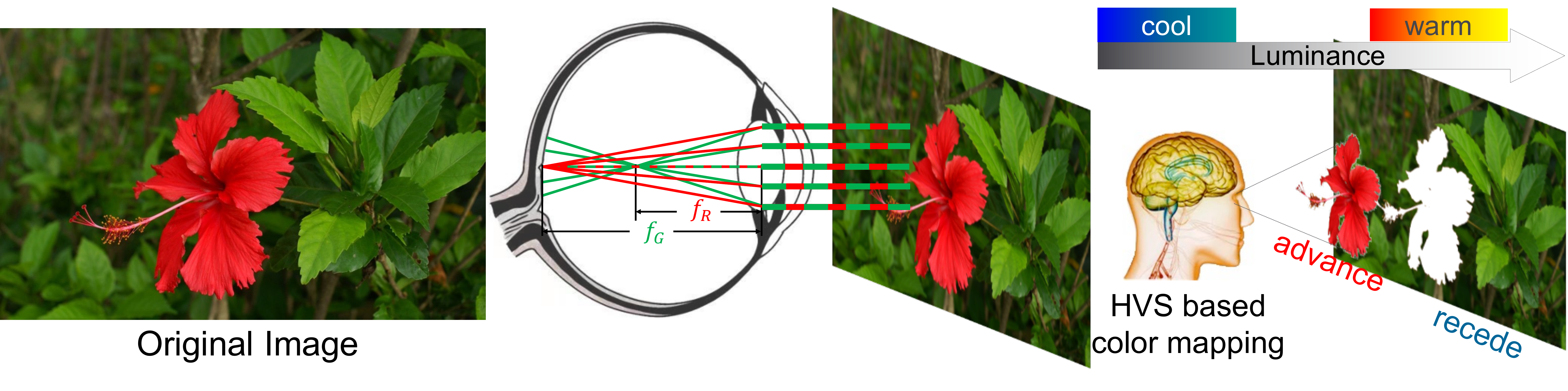}
	\caption{Chromatic aberration which results from differential refraction of light depending on its wavelength, it causes some rays (green) to converge before other (red). This results in a perception of red ``advancing'' toward the eye, while green to be ``receding''.}   
	\label{fig:t2}
\end{figure*}

\begin{figure*}[!t]
	\centering
	\includegraphics[width = \textwidth]{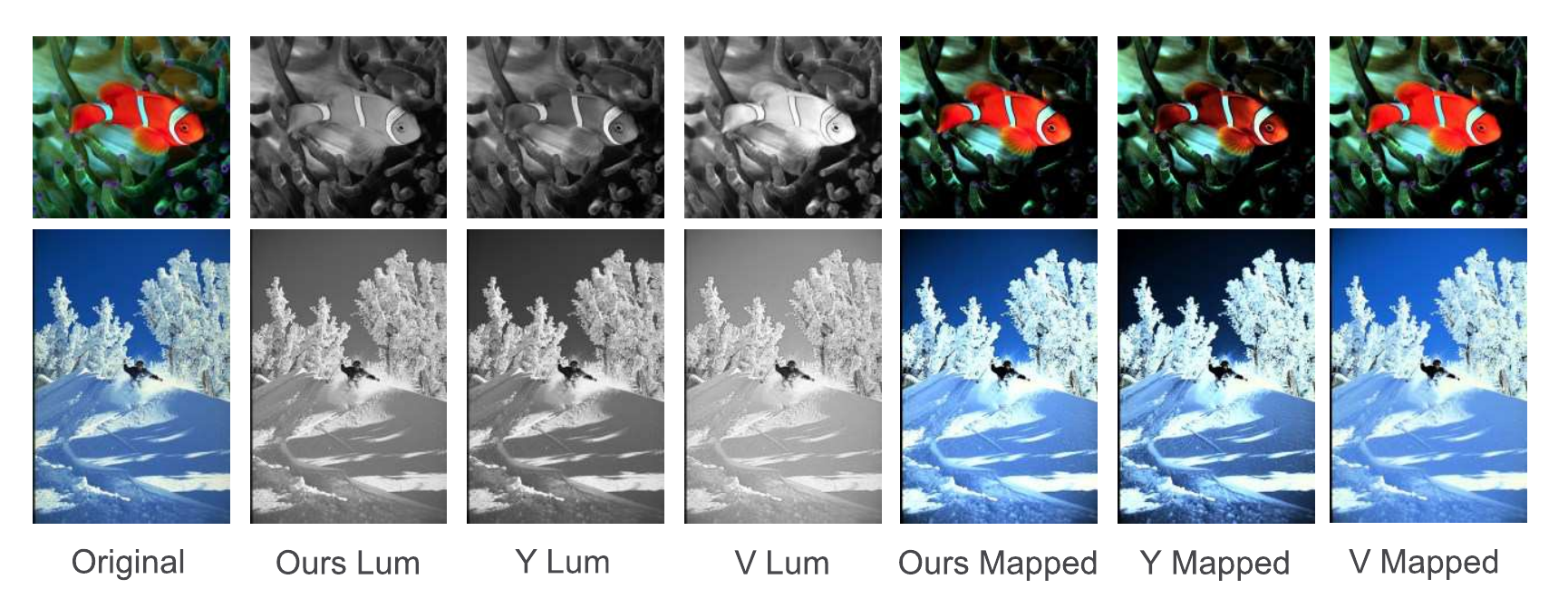}
	\caption{Global tone mapping performed in three different luminance spaces: Our's, Y of YCbCr and V of HSV}
	\label{fig:GTMO}
\end{figure*}

\begin{table}[!t]
\scriptsize
	\renewcommand{\arraystretch}{1.2}
	\caption{Comparison of six color to gray methods based on average objective metrics for 250 images.}
		\centering
	\begin{tabular}{| l | l | l | l | l | l |}
		\hline
		& Decolorization & CCFR & CCPR & E-score  & C2G-SSIM \\ \hline
		(a) & Lu et al. \cite{lu2014contrast} & 0.9922 & 0.9645 & 0.9781 & 0.8900 \\ \hline
		(b) & Nafchi et al. \cite{nafchi2017corrc2g} & 0.9880 & 0.9555 & 0.9710 & 0.8900 \\ \hline
		(c) & Grundland et al \cite{grundland2007decolorize} & 0.9811 & 0.9584 & 0.9747 & 0.8749 \\ \hline
		(d) & Smith et al. \cite{smith2008apparent} & 0.9880 & 0.9555 & 0.9710 & 0.9062 \\ \hline
		(e) & Gooch et al. \cite{gooch2005color2gray} & 0.9839 & 0.9545 & 0.9714 & 0.8935 \\ \hline   
		(f) & Ours & 0.9890 & 0.9576 & 0.9728 & 0.9018 \\ \hline
	\end{tabular}
	\label{tab_om}
\end{table}

In our experiment we computed the average CCPR for the 250 images in the dataset by varying $\tau$ from 1 to 15 \cite{lu2014contrast}. As can be seen from Fig. \ref{fig:plotE}, our algorithm's performance is reasonable and practicable when compared to other color to gray algorithms. Figure \ref{fig:ssim} shows the average C2G-SSIM score for the six decolorization methods. According to the plot in Fig. \ref{fig:plotE}, Lu et al.'s method shows best performance based on the E-score, however, our method delivers high average C2G-SSIM measure which is better correlated to human perception \cite{nafchi2017corrc2g}. 

\section{Characteristics of visualization of our algorithm}
\subsection{Color temperature and chromatic aberration}\label{ca}
From Snell's law we know that the refraction of light is dependent on its wavelength. As the frequency of light increases, its refractive index becomes larger, causing more refraction of the shorter wavelengths. Therefore, when an image is captured through a lens, all colors do not focus at the same distance, and these imperfections are known as chromatic aberration. In cameras this imperfection is removed by using a combination of second achromatic lens which is made of different material(glass) than the first lens. This second lens would reverse the color dispersion caused by the first lens. The human eye, also employs a lens and does exhibit this phenomena as shown in Fig. \ref{fig:t2}. From this figure we can observe that red light forms the image farthest from the lens as it has the smallest refractive index. Colors with higher index of refraction would ideally bend more thereby forming images closer to the lens. Therefore, it would be impossible to focus on all colors simultaneously, resulting in ``somewhat fuzzy'' images that are not in focus. Colors that are closer to red end of the electromagnetic spectrum are said to be warm colors and are perceived as closer to the observer \cite{niemeyer1988ec88,eastlake2020goethe}. Colors that are around the blue end are said to be cooler colors and are perceived to be receding away from the observer. This phenomenon has been exploited by traditional artists to add depth information in artwork \cite{nguyen2005detection}, display devices \cite{hodges1988chromostereoscopic}, and 3D imagery \cite{hong2012depth}. 

\begin{figure*}[!t]
	\centering
	\includegraphics[width = \textwidth]{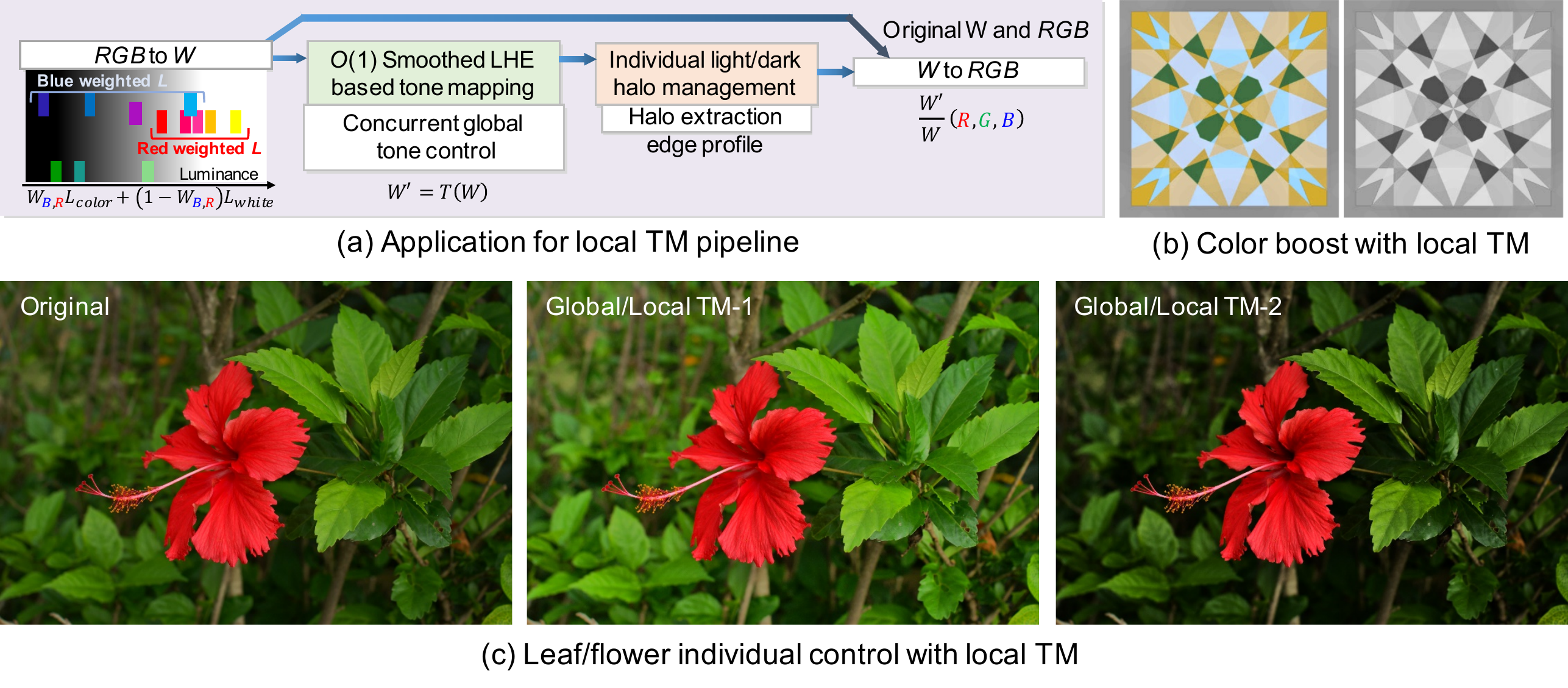}
	\caption{Application: (a) Ideal image processing pipeline (tone mapping) (b) Color boosted by local tone mapping and corresponding grayscale image. (c) Demonstration example for easy control features using our decolorization algorithm. TM1: background boost TM2: background suppression}
	\label{fig:SLHE_TMO}
\end{figure*}

\subsection{Tone mapping application}
The color to grayscale conversion is a dimensionality reduction problem whose significance has been underestimated. Usually the tone mapping is performed on the grayscale image because of the lower computational and memory requirements, when compared to tone mapping on the RGB channels. In this section we will discuss global and local tone mapping application using the proposed decolorization algorithm and demonstrate its effectiveness.   
\subsubsection{Global tone mapping}

We discuss our decolorization method's applicability for post-processing tasks such as tone mapping. Conventional methods have not shown such applicability because of their long run-time; thus, the use of $Y$ of YCbCr or $V$ of HSV for tone mapping is popular today. We evaluated our method's output of global tone mapping because it has the ability to adequately remap colors to $1-D$ luminance. Thus, even using global tone mapping confirms the effectiveness of our method. Figure \ref{fig:GTMO} shows tone mapped results using $Y$, $V$, or our luminance channel. In the fish image, using $Y$, the fish color became the same as the background and could not be controlled separately, because the weight of $R$ is small in the YCbCr color channel. Our method was able to separate fish color from background in the same way as human perception and directly mapped the color of the background to dark. In the sky and snow scene, the global tone curve for contrast enhancement is a centered sigmoid curve, in which highlight/shadow clippings occur naturally. Since HSV color space treats primary colors and whites the same way, colors of sky and snow were mapped to luminance that was too light. Thus, the sigmoid function degraded the contrast of the outputted image. YCbCr treats $B$ as too low luminance; the sky became too dark. We confirmed that our method generates well-balanced images maintaining the contrast and the colors. 

\subsubsection{Local tone mapping}
Local histogram equalization based local tone mapping converts target pixels by using tone curves constructed from local cumulative histograms. Smoothed local histogram equalization (LHE) are also used as a smoothed LH filter \cite{kass2010smoothed}. The Apical's (ARM\textsuperscript{\textregistered}) Iridix algorithm \cite{chesnokov2007image}, which is based on smoothed LHE are used by a range of camera makers, including Nikon, Olympus and Sony. For our local tonemap application we selected smoothed LHE-based function as they are very suitable for practical real-time applications, in Fig. \ref{fig:SLHE_TMO}(a) we show an ideal pipeline for such a system. We implemented the local tonemap application similar to one presented by Ambalathankandy et al. \cite{ambalathankandy2019adaptive}, as their implementation has a linear $\textit{O}(1)$ computational complexity and produces output images with fine quality as shown in Fig. \ref{fig:SLHE_TMO}(b, c). The total computational time including our proposed decolorization and the local tonemap operation was only $14.7\mu s$ per pixel (normalized time $@2.7$GHz CPU). This time utilization is $20\times$ less than Lu et al.'s work \cite{lu2012contrast}. Additionally, using our decolorization method has an advantage, which is to individually control the back/foreground as shown in Fig. \ref{fig:SLHE_TMO}(c). 

\subsection{Limitation of our method}
\begin{figure*}[!t]
	\centering
	\includegraphics[width = \textwidth]{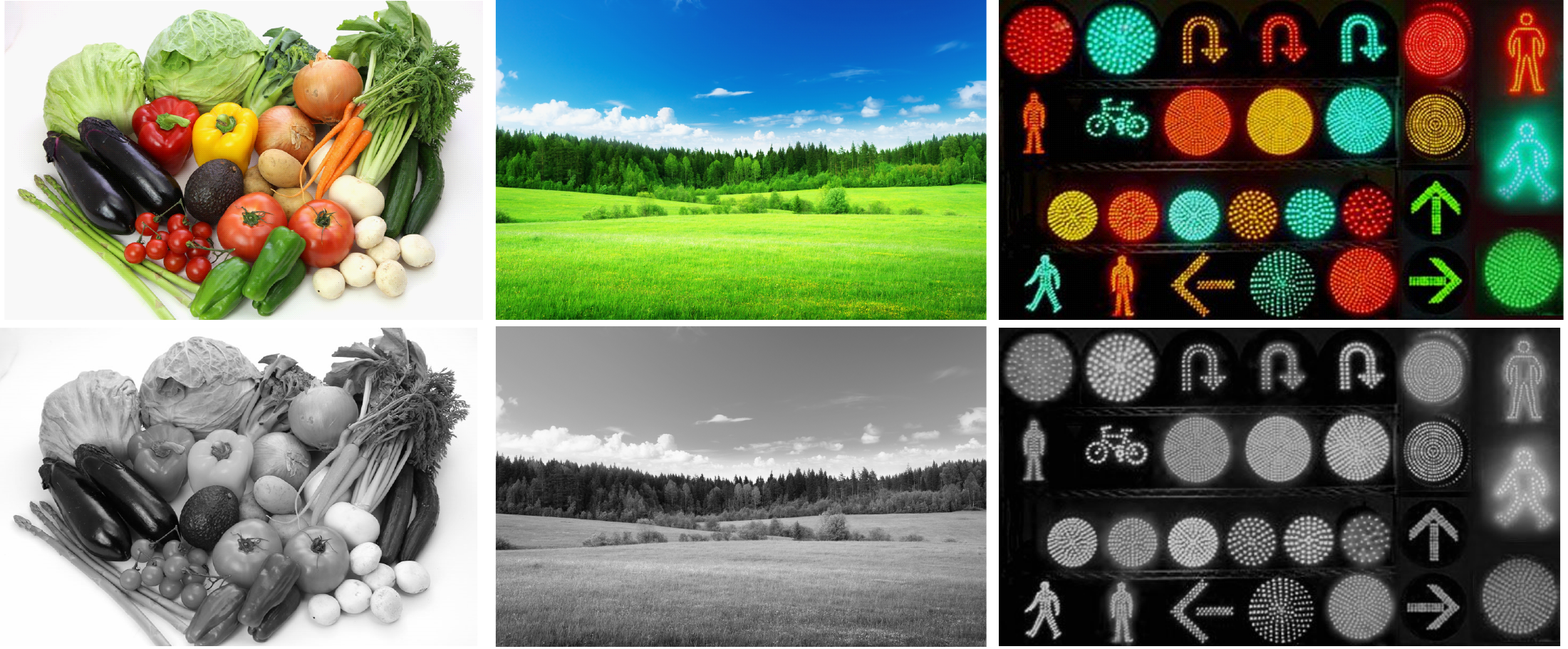}
	\caption{Examples to demonstrate limitation of our decolorization method when mapping bright green/neon color.}
	\label{fig:Limit}
\end{figure*}

In our decolorization method, pure green is likely to be mapped to a dark luminance value. Light green will also be mapped to a lesser dark luminance part as shown in Fig. \ref{fig:4}. Therefore, certain scenes are likely to be perceived as unnatural. For example: 
(a) Vegetables (e.g. leafy greens like cabbage) (b) Green meadows under bright sky. (c) Bright green neon lights. However, in Fig. \ref{fig:Limit} we perceive them as natural. We postulate the following as the possible reasons: (i) There are rarely any pure bright green 
(like G255) scene in nature. (ii) In our color space, green color with white components follow Eq.(\ref{eq:Lw1}) by weighting function. Therefore, the color keeps a balance among other color channels. (iii) Vegetation scenery with dark green are well perceived as healthy plants.

\section{Conclusion}

In this paper, we present a color temperature-based RGB to gray conversion model by taking in to account the chromatic aberration phenomena. This anomaly results from differential refraction of light depending on its wavelength, it causes some of the rays (cool colors) to converge before others (warm colors). This results in a perception of warmer colors ``advancing'' towards the eye, while the cooler ones to be ``receding''. Essentially, since decolorization is expected to have a key role in the pre-processing of tone mapping or edge preserving filters, low calculation cost and fast operation for the processing are required. To address this requirement, we have developed a high-speed $\textit{O}(1)$ decolorization method that is based on color temperature-based perception. It refers to RGB values in one pixel and performs weighted blending of the Euclidean distances of warm/cool color vectors. 
This simple conversion outputs a gray channel that is comparable to the conventional optimization methods using iterations. When our method is applied to tone mapping, it achieves better results than one could be obtained with YCbCr/HSV color space.

\section{Acknowledgments}
We would like to sincerely thank Cadik et al., Lu et al. for making available their datasets, and other authors for sharing their source codes.

\bibliographystyle{plain}

\bibliography{report}

\begin{thebibliography}{10}
\providecommand{\url}[1]{#1}
\csname url@samestyle\endcsname
\providecommand{\newblock}{\relax}
\providecommand{\bibinfo}[2]{#2}
\providecommand{\BIBentrySTDinterwordspacing}{\spaceskip=0pt\relax}
\providecommand{\BIBentryALTinterwordstretchfactor}{4}
\providecommand{\BIBentryALTinterwordspacing}{\spaceskip=\fontdimen2\font plus
\BIBentryALTinterwordstretchfactor\fontdimen3\font minus
  \fontdimen4\font\relax}
\providecommand{\BIBforeignlanguage}[2]{{%
\expandafter\ifx\csname l@#1\endcsname\relax
\typeout{** WARNING: IEEEtran.bst: No hyphenation pattern has been}%
\typeout{** loaded for the language `#1'. Using the pattern for}%
\typeout{** the default language instead.}%
\else
\language=\csname l@#1\endcsname
\fi
#2}}
\providecommand{\BIBdecl}{\relax}
\BIBdecl

\bibitem{morton1998color}
J.~L. Morton, \emph{Color logic}.\hskip 1em plus 0.5em minus 0.4em\relax
  Colorcom, 1998.

\bibitem{sundet1978effects}
J.~M. SUNDET, ``Effects of colour on perceived depth: review of experiments and
  evalutaion of theories,'' \emph{Scandinavian Journal of Psychology}, vol.~19,
  no.~1, pp. 133--143, 1978.

\bibitem{Nayatani1}
Y.~Nayatani, ``Simple estimation methods for the helmholtz kohlrausch effect,''
  \emph{Color Research \& Application}, vol.~22, no.~6, pp. 385--401, 1997.

\bibitem{gooch2005color2gray}
A.~A. Gooch, S.~C. Olsen, J.~Tumblin, and B.~Gooch, ``Color2gray:
  salience-preserving color removal,'' \emph{ACM Transactions on Graphics
  (TOG)}, vol.~24, no.~3, pp. 634--639, 2005.

\bibitem{grundland2007decolorize}
M.~Grundland and N.~A. Dodgson, ``Decolorize: Fast, contrast enhancing, color
  to grayscale conversion,'' \emph{Pattern Recognition}, vol.~40, no.~11, pp.
  2891--2896, 2007.

\bibitem{smith2008apparent}
K.~Smith, P.-E. Landes, J.~Thollot, and K.~Myszkowski, ``Apparent greyscale: A
  simple and fast conversion to perceptually accurate images and video,'' in
  \emph{Computer Graphics Forum}, vol.~27, no.~2, pp. 193--200.\hskip 1em plus
  0.5em minus 0.4em\relax Wiley Online Library, 2008.

\bibitem{kim2009robust}
Y.~Kim, C.~Jang, J.~Demouth, and S.~Lee, ``Robust color-to-gray via nonlinear
  global mapping,'' \emph{ACM Transactions on Graphics (TOG)}, vol.~28, no.~5,
  p. 161, 2009.

\bibitem{lu2012contrast}
C.~Lu, L.~Xu, and J.~Jia, ``Contrast preserving decolorization,'' in
  \emph{Computational Photography (ICCP), 2012 IEEE International Conference
  on}, pp. 1--7.\hskip 1em plus 0.5em minus 0.4em\relax IEEE, 2012.

\bibitem{ancuti2010image}
C.~O. Ancuti, C.~Ancuti, C.~Hermans, and P.~Bekaert, ``Image and video
  decolorization by fusion,'' in \emph{Asian Conference on Computer Vision},
  pp. 79--92.\hskip 1em plus 0.5em minus 0.4em\relax Springer, 2010.

\bibitem{ancuti2016laplacian}
C.~Ancuti and C.~O. Ancuti, ``Laplacian-guided image decolorization,'' in
  \emph{Image Processing (ICIP), 2016 IEEE International Conference on}, pp.
  4107--4111.\hskip 1em plus 0.5em minus 0.4em\relax IEEE, 2016.

\bibitem{liu2015gcsdecolor}
Q.~Liu, P.~X. Liu, W.~Xie, Y.~Wang, and D.~Liang, ``Gcsdecolor: gradient
  correlation similarity for efficient contrast preserving decolorization,''
  \emph{IEEE Transactions on Image Processing}, vol.~24, no.~9, pp. 2889--2904,
  2015.

\bibitem{song20161}
Y.~Song and L.~Gong, ``O (1) contrast preserving decolorization using linear
  local mapping,'' in \emph{Wireless Communications \& Signal Processing
  (WCSP), 2016 8th International Conference on}, pp. 1--6.\hskip 1em plus 0.5em
  minus 0.4em\relax IEEE, 2016.

\bibitem{liu2019image}
S.~Liu and X.~Zhang, ``Image decolorization combining local features and
  exposure features,'' \emph{IEEE Transactions on Multimedia}, 2019.

\bibitem{cai2018perception}
B.~Cai, X.~Xu, and X.~Xing, ``Perception preserving decolorization,'' in
  \emph{2018 25th IEEE International Conference on Image Processing (ICIP)},
  pp. 2810--2814.\hskip 1em plus 0.5em minus 0.4em\relax IEEE, 2018.

\bibitem{hou2017deep}
X.~Hou, J.~Duan, and G.~Qiu, ``Deep feature consistent deep image
  transformations: Downscaling, decolorization and hdr tone mapping,''
  \emph{arXiv preprint arXiv:1707.09482}, 2017.

\bibitem{lin2008learning}
Z.~Lin, W.~Zhang, and X.~Tang, ``Learning partial differential equations for
  computer vision,'' \emph{Peking Univ., Chin. Univ. of Hong Kong}, 2008.

\bibitem{zhang2018contrast}
X.~Zhang and S.~Liu, ``Contrast preserving image decolorization combining
  global features and local semantic features,'' \emph{The Visual Computer},
  vol.~34, no. 6-8, pp. 1099--1108, 2018.

\bibitem{simonyan2018very}
K.~Simonyan and A.~Zisserman, ``Very deep convolutional networks for
  large-scale image recognition. 2014,'' \emph{arXiv preprint arXiv:1409.1556},
  2018.

\bibitem{ou2004study}
L.-C. Ou, M.~R. Luo, A.~Woodcock, and A.~Wright, ``A study of colour emotion
  and colour preference. part i: Colour emotions for single colours,''
  \emph{Color Research \& Application}, vol.~29, no.~3, pp. 232--240, 2004.

\bibitem{cadik2008perceptual}
M.~{\^C}ad{\'\i}k, ``Perceptual evaluation of color-to-grayscale image
  conversions,'' in \emph{Computer Graphics Forum}, vol.~27, no.~7, pp.
  1745--1754.\hskip 1em plus 0.5em minus 0.4em\relax Wiley Online Library,
  2008.

\bibitem{shan2009globally}
Q.~Shan, J.~Jia, and M.~S. Brown, ``Globally optimized linear windowed tone
  mapping,'' \emph{IEEE transactions on visualization and computer graphics},
  vol.~16, no.~4, pp. 663--675, 2009.

\bibitem{lu2014contrast}
C.~Lu, L.~Xu, and J.~Jia, ``Contrast preserving decolorization with
  perception-based quality metrics,'' \emph{International journal of computer
  vision}, vol. 110, no.~2, pp. 222--239, 2014.

\bibitem{ma2015objective}
K.~Ma, T.~Zhao, K.~Zeng, and Z.~Wang, ``Objective quality assessment for
  color-to-gray image conversion,'' \emph{IEEE Transactions on Image
  Processing}, vol.~24, no.~12, pp. 4673--4685, 2015.

\bibitem{wang2004image}
Z.~Wang, A.~C. Bovik, H.~R. Sheikh, and E.~P. Simoncelli, ``Image quality
  assessment: from error visibility to structural similarity,'' \emph{IEEE
  transactions on image processing}, vol.~13, no.~4, pp. 600--612, 2004.

\bibitem{nafchi2017corrc2g}
H.~Z. Nafchi, A.~Shahkolaei, R.~Hedjam, and M.~Cheriet, ``Corrc2g: Color to
  gray conversion by correlation,'' \emph{IEEE Signal Processing Letters},
  vol.~24, no.~11, pp. 1651--1655, 2017.

\bibitem{niemeyer1988ec88}
S.~Niemeyer, ``Ec88-423 color expression in the home,'' 1988.

\bibitem{eastlake2020goethe}
C.~L. Eastlake, \emph{Goethe's Theory of Colours}.\hskip 1em plus 0.5em minus
  0.4em\relax BoD--Books on Demand, 2020.

\bibitem{nguyen2005detection}
V.~A. Nguyen, I.~P. Howard, and R.~S. Allison, ``Detection of the depth order
  of defocused images,'' \emph{Vision Research}, vol.~45, no.~8, pp.
  1003--1011, 2005.

\bibitem{hodges1988chromostereoscopic}
L.~F. Hodges and D.~F. McAllister, ``Chromostereoscopic crt-based display,'' in
  \emph{Three-Dimensional Imaging and Remote Sensing Imaging}, vol. 902, pp.
  37--45.\hskip 1em plus 0.5em minus 0.4em\relax International Society for
  Optics and Photonics, 1988.

\bibitem{hong2012depth}
J.~Hong, H.~Lee, D.~Park, and C.~Kim, ``Depth perception enhancement based on
  chromostereopsis in a 3d display,'' \emph{Journal of Information Display},
  vol.~13, no.~3, pp. 101--106, 2012.

\bibitem{kass2010smoothed}
M.~Kass and J.~Solomon, ``Smoothed local histogram filters,'' \emph{ACM
  Transactions on Graphics (TOG)}, vol.~29, no.~4, p. 100, 2010.

\bibitem{chesnokov2007image}
V.~Chesnokov, ``Image enhancement methods and apparatus therefor,'' Nov.~27
  2007, uS Patent 7,302,110.

\bibitem{ambalathankandy2019adaptive}
P.~Ambalathankandy, M.~Ikebe, T.~Yoshida, T.~Shimada, S.~Takamaeda,
  M.~Motomura, and T.~Asai, ``An adaptive global and local tone mapping
  algorithm implemented on fpga,'' \emph{IEEE Transactions on Circuits and
  Systems for Video Technology}, 2019.

\end{thebibliography}

\end{document}